\documentclass[letter]{llncs}

\usepackage[LSBC1,T1]{fontenc}
\usepackage[latin1]{inputenc}

\usepackage{chessfss}
\setboardfontsize{8pt}
\setboardfontencoding{LSBC1}
\setboardfontcolors{
blackonwhitepiecemask=white, blackpiece=black}
\newcommand{\brook}{\raisebox{-8pt}{\WhiteRookOnWhite}}

\usepackage{amsmath}
\usepackage{amssymb}
\usepackage{sidecap}
\usepackage{wrapfig}
\usepackage{rotating}

\usepackage{tikz}
\usetikzlibrary{matrix}
\usetikzlibrary{decorations.pathreplacing}
\usetikzlibrary{mindmap}
\usetikzlibrary{calc}
\usetikzlibrary{positioning}
\usetikzlibrary{chains}
\definecolor{rred}{RGB}{150,35,31}
\tikzstyle{every label}=[rred,font=\sf]
\tikzstyle{every pin}=[rred,font=\sf, inner sep=1pt]
\tikzstyle{every pin edge}=[rred,<-,>=stealth,shorten <=2pt]
\tikzstyle{vertex}=[inner sep= 1pt,draw,circle,font=\sf\small]
\definecolor{lblue}{RGB}{176,196,222}
\newcommand{\lblue}{lblue}
\definecolor{dblue}{RGB}{70, 130, 180}

\usepackage{sidecap}

\renewcommand{\emptyset}{\varnothing}

\newenvironment{pf}[1][Proof]{\medskip\noindent\emph{#1. }}{\qed\smallskip}

\gdef\GITAbrHash{dc9a128}\gdef\GITAuthorDate{Tue May 17 14:09:02 2011 +0200}\gdef\GITAuthorName{Thore Husfeldt}

\makeatletter

\renewcommand\section{\@startsection {section}{2}{0pt} {-3.5ex \@plus -1ex \@minus -.2ex}{-2.3ex \@plus.2ex}{\bfseries}}

\title{Invitation to Algorithmic Uses of Inclusion--Exclusion}

\author{Thore Husfeldt}
\institute{IT University of Copenhagen, Denmark\\Lund University, Sweden}

\begin{document}

\maketitle
\let\thefootnote\relax
\footnotetext{Base revision~\GITAbrHash$\ldots$, \GITAuthorDate, \GITAuthorName.}
\begin{abstract}
  I give an introduction to algorithmic uses of the principle of
  inclusion--exclusion. %
  The presentation is intended to be be concrete and accessible, at
  the expense of generality and comprehensiveness.
\end{abstract}

\tikzstyle{vertex}=[inner sep= 1pt,draw,circle,font=\sf\small] 
\newcommand{\hasselines}{(0) -- (1) (0) -- (2) (0) -- (3) (0) -- (4)
  (1) -- (12) (1) -- (13) (1) -- (14) (2) -- (12) (2) -- (23) (2) --
  (24) (3) -- (13) (3) -- (23) (3) -- (34) (4) -- (14) (4) -- (24)
  (4) -- (34) (12) -- (123) (13)--(123) (23)--(123) (12)--(124)
  (14)--(124) (24) -- (124) (13)--(134) (14)--(134) (34)--(134)
  (23)--(234) (34)--(234) (24)--(234) (123)--(1234) (124)--(1234)
  (134)--(1234) (234)--(1234)
}
\newcommand{\drawhassenodes}{
  \node [vertex] (0)    at (0,0)  {};
  \node [vertex] (1)    at (-1,1) {};
  \node [vertex] (2)    at (0,1)  {};
  \node [vertex] (3)    at (1,1)  {};
  \node [vertex] (4)    at (3,1)  {};
  \node [vertex] (12)   at (-1,2) {};
  \node [vertex] (13)   at (0,2)  {};
  \node [vertex] (23)   at (1,2)  {};
  \node [vertex] (14)   at (2,2)  {};
  \node [vertex] (24)   at (3,2)  {};
  \node [vertex] (34)   at (4,2)  {};
  \node [vertex] (123)  at (0,3)  {};
  \node [vertex] (124)  at (2,3)  {};
  \node [vertex] (134)  at (3,3)  {};
  \node [vertex] (234)  at (4,3)  {};
  \node [vertex] (1234) at (3,4)  {};
}

\begin{wrapfigure}{r}{3.2cm}
\vspace{-4ex}
\begin{tikzpicture}[scale=.5]
  \drawhassenodes
  \node [fill=rred, draw, circle, very thick,inner sep=2pt,
  label=below left: \small $R$] (1) at (-1,1) {};
  \node [fill=rred, draw, circle, very thick,inner sep=2pt, pin=
  above left: \small $T$] (124) at (2,3) {};
  \node [fill=lblue, draw, circle, very thick,inner sep=2pt] (12) at (-1,2) {};
  \node [fill=lblue, draw, circle, very thick,inner sep=2pt] (14) at (2,2) {};
  \draw \hasselines;
\end{tikzpicture}
\end{wrapfigure}

\section{The principle of inclusion--exclusion.}

There are as many odd-sized as even-sized subsets sandwiched between
two different sets: For $R\subseteq T$,
\begin{equation}\label{eq: pie}
\sum_{R\subseteq S\subseteq T}  (-1)^{|T\setminus S|} = [R=T]\,.
\end{equation}

We use Iverson notation $[P]$ for proposition $P$, meaning $[P]=1$ if
$P$ and $[P]=0$ otherwise.

\begin{pf}[Proof of \eqref{eq: pie}]
  If $R=T$ then there is exactly one sandwiched set, namely $S=T$. %
  Otherwise we set up a bijection between the odd- and even-sized
  subsets as follows. %
  Fix $t\in T\setminus R$. %
  For every odd-sized subset $S_1$ with $R\subseteq S_1\subseteq T$
  let $S_0= S_1\oplus \{t\}$ denote the symmetric difference of $S_1$
  with $\{t\}$. %
  Note that the size of $S_0$ is even and that $S_0$ contains $R$. %
  Furthermore, $S_1$ can be recovered from $S_0$ as $S_1=S_0\oplus
  \{t\}$. %
\end{pf}

\paragraph{Perspective.} 
We will see the (perhaps more familiar) formulation of the principle
of inclusion--exclusion in terms of intersecting sets in
\S\ref{sec: ie sets}, %
and another equivalent formulation in \S\ref{sec:
  mobin}.

\begin{wrapfigure}{r}{2.2cm}
  \vspace{-4ex}
  \begin{tikzpicture}[scale=.5]
    \filldraw[lblue] (0:1)  circle  (.75);
    \filldraw[lblue] (240:1) circle (.75);
    \filldraw[lblue] (120:1) circle (.75);
    \filldraw[lblue] (60:1.8) circle (.75);
    \fill [fill=lblue, decorate,decoration={start
      radius=.75cm,end radius=.75cm,amplitude=10,angle=30, circle
      connection bar}] (120:1) -- (60:1.8);
    \begin{scope}[every node/.style={circle, fill=white,draw,thick,inner sep=1pt,font=\sf\small}]
      \node (1) at (0:1) {A};
      \node (2) at (120:1) {B};    
      \node (3) at (240:1) {C};
      \node (4) at (60:1.8) {D};
      \draw [thick](4)--(1)--(2)--(3)--(1);
    \end{scope}
    \node at (0:2.2) {\small $I_1$};
    \node at (240:2.2) {\small $I_2$};
    \node at (100:2.2) {\small $I_3$};
  \end{tikzpicture}
\end{wrapfigure}

\section{Graph colouring.}
\label{sec: colouring}

A $k$-\emph{colouring} of a graph $G=(N,E)$ on $n=|N|$ nodes assigns
one of $k$ colours to every node such that neighbouring nodes have
different colours. %
In any such colouring, the nodes of the same colour form a nonempty
\emph{independent set}, a set of nodes none of which are neighbours.

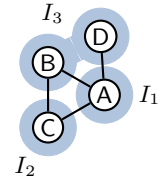
\newcommand{\ind}{g}

Let $\ind(S)$ denote the number of nonempty independent subsets in
$S\subseteq N$. %
Then $G$ can be $k$-coloured if and only if %
\begin{equation}\label{eq: ie c} 
  \sum_{S\subseteq N}
  (-1)^{n-|S|} \bigl(\ind(S)\bigr)^k > 0 \,.
\end{equation}

\begin{pf}
  For every $S\subseteq N$, the term $\ind(S)^k$ counts the number of
  ways to pick $k$ nonempty independent sets $I_1,\ldots, I_k$ in $S$. %
  Thus, we can express the left hand side of \eqref{eq: ie c} as 
  \[ \sum_S \sum_{I_1} \cdots  \sum_{I_k} [\,\forall
  i\colon I_i\subseteq S\,]
  (-1)^{|N\setminus S|}  = 
  \sum_{I_1} \cdots  \sum_{I_k}
  \sum_{S} [\,\forall i\colon I_i\subseteq S\,]
  (-1)^{|N\setminus S|}\,.
  \]
  The innermost sum has the form
  \[
  \sum_{I_1\cup \cdots\cup I_k\subseteq S\subseteq N} (-1)^{|N\setminus S|}\,.
  \]
  By \eqref{eq: pie}, the only contributions come from $I_1\cup \cdots\cup I_k=N$.
  Every such choice indeed corresponds to a valid colouring: %
  For $i=1,\ldots,k$, let the nodes in $I_i$ have colour $i$. %
  (This may re-colour some nodes.) %
  Conversely, every valid $k$-colouring corresponds to such a choice.
  (In fact, the colourings are the disjoint partitions).
\end{pf}

\begin{figure}
  \[
  \vcenter{\hbox{
  \begin{tikzpicture}[scale=.5]
    \begin{scope}[every node/.style={circle, fill=white,draw,thick,inner sep=1pt,font=\sf}]
      \node (1) at (0:1) {A};
      \node (2) at (120:1) {B};    
      \node (3) at (240:1) {C};
      \node (4) at (60:1.8) {D};
      \draw [thick](4)--(1)--(2)--(3)--(1);
    \end{scope}
  \end{tikzpicture}}}
  \qquad
  \tikzstyle{vertex}=[inner sep= 1pt,draw,circle,font=\sf\small]
  \vcenter{\hbox{
  \begin{tikzpicture}[xscale=1.25,every label/.style={rred,font=\sf}]
    \node [vertex] (0)    at (0,0)  {0};
    \node [vertex, label = right:A] (1)    at (-1,1) {1};
    \node [vertex, label = right:B] (2)    at (0,1)  {1};
    \node [vertex, label =  right:C] (3)    at (1,1)  {1};
    \node [vertex, label =  right:D] (4)    at (3,1)  {1};
    \node [vertex, label = right:AB] (12)   at (-1,2) {2};
    \node [vertex, label = right:AC] (13)   at (0,2)  {2};
    \node [vertex, label = right:BC] (23)   at (1,2)  {2};
    \node [vertex, label = right:AD] (14)   at (2,2)  {2};
    \node [vertex, label = right:BD] (24)   at (3,2)  {3};
    \node [vertex, label = right:CD] (34)   at (4,2)  {3};
    \node [vertex, label = left:ABC] (123)  at (0,3)  {3};
    \node [vertex, label = left:ABD] (124)  at (2,3)  {4};
    \node [vertex, label = right:ACD] (134)  at (3,3)  {4};
    \node [vertex, label = right:BCD] (234)  at (4,3)  {5};
    \node [vertex, label = right:ABCD] (1234) at (3,4)  {6};
    \node [rred] (legend) at (0,4) {\small $g(\{\mathsf A,\mathsf
      C,\mathsf D\})=4$};
    \draw [rred,->] (legend) -- (134);
    \draw \hasselines;
  \end{tikzpicture}
}}
  \]
  \caption{%
    The values of $\ind(S)$ for all $S$ for the example graph to the
    left. %
    Expression \eqref{eq: ie c} evaluates an alternating sum of
    the cubes of these values, in this case $\binoppenalty=0 6^3
    -(3^3+4^3+4^3+5^3) + (2^3+2^3+2^3+2^3+3^3+3^3) - (1^3 + 1^3 + 1^3 + 1^3) + 0 = 18$. }
\end{figure}
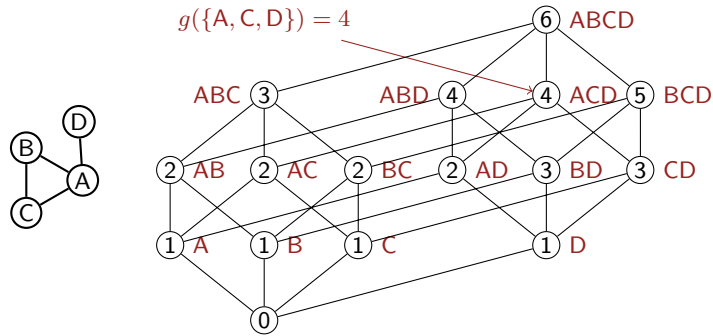

\section{Counting the number of independent sets.}
\label{sec: compute g}
Expression \eqref{eq: ie c} can be evaluated in two ways:

For each $S\subseteq N$, the value $\ind(S)$ can be computed in time
$O(2^{|S|}|E|)$ by constructing every nonempty subset of $S$ and testing it
for independence. %
Thus, the total running time for evaluating $\eqref{eq: ie c}$ is
within a polynomial factor of
\[  \sum_{S\subseteq N} 2^{|S|} = \sum_{i=1}^n \binom{n}{i} 2^i = 3^n\,.\]
The space requirement is polynomial.

Alternatively, we first build a table with $2^n$ entries containing
$\ind(S)$ for all $S\subseteq N$, after which we can evaluate \eqref{eq: ie
  c} in time and space $2^nn^{O(1)}$.

Such a table is easy to build given a recurrence for $\ind(S)$. We have
$\ind(\emptyset)=0$, and
\begin{equation}\label{eq: ind(S)}
  \ind\bigl(S\bigr)= \ind\bigl(S\setminus \{v\}\bigr) +
  \ind\bigl(S\setminus  N[v]\bigr)+1\qquad (v\in S)\,,
\end{equation}
where $N[v]=\{v\}\cup\{\,u\in N\colon uv\in E\,\}$ denotes the closed
neighbourhood of $v$.

\begin{pf}[Proof of \eqref{eq: ind(S)}] %
  Fix $v\in S$ and consider the nonempty independent sets $I\subseteq S$. %
  They can be partitioned into two classes: either $v\in I$ or
  $v\notin I$. %
  The latter sets are counted in $\ind\bigl(S\setminus\{v\}\bigr)$. %
  It remains to argue that the sets $I\ni v$ are counted in
  $\ind\bigl(S\setminus N[v]\bigr)+1$. %
  We will do this by counting the equipotent family of sets $I\setminus\{v\}$
  instead. %
  Since $I$ contains $v$ and is independent, it cannot contain other
  nodes in $N[v]$. %
  Thus $I\setminus\{v\}$ is disjoint from $N[v]$ and contained in $S$. %
  Now, either $I$ is the singleton $\{v\}$ itself, accounted for by
  the `$+1$' term, or $I\setminus\{v\}$ is a nonempty independent set and
  therefore counted in $\ind\bigl(S\setminus N[v]\bigr)$.
\end{pf}

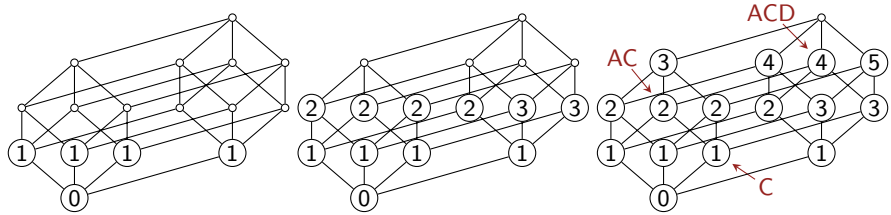
\begin{figure}
     \begin{tikzpicture}[yscale=.6,xscale=.7]
    \node [vertex] (0)    at (0,0)  {0};
    \node [vertex] (1)    at (-1,1) {1};
    \node [vertex] (2)    at (0,1)  {1};
    \node [vertex] (3)    at (1,1)  {1};
    \node [vertex] (4)    at (3,1)  {1};
    \node [vertex] (12)   at (-1,2) {};
    \node [vertex] (13)   at (0,2)  {};
    \node [vertex] (23)   at (1,2)  {};
    \node [vertex] (14)   at (2,2)  {};
    \node [vertex] (24)   at (3,2)  {};
    \node [vertex] (34)   at (4,2)  {};
    \node [vertex] (123)  at (0,3)  {};
    \node [vertex] (124)  at (2,3)  {};
    \node [vertex] (134)  at (3,3)  {};
    \node [vertex] (234)  at (4,3)  {};
    \node [vertex] (1234) at (3,4)  {};
    \draw \hasselines ;
  \end{tikzpicture}
 \begin{tikzpicture}[yscale=.6,xscale=.7]
    \node [vertex] (0)    at (0,0)  {0};
    \node [vertex] (1)    at (-1,1) {1};
    \node [vertex] (2)    at (0,1)  {1};
    \node [vertex] (3)    at (1,1)  {1};
    \node [vertex] (4)    at (3,1)  {1};
    \node [vertex] (12)   at (-1,2) {2};
    \node [vertex] (13)   at (0,2)  {2};
    \node [vertex] (23)   at (1,2)  {2};
    \node [vertex] (14)   at (2,2)  {2};
    \node [vertex] (24)   at (3,2)  {3};
    \node [vertex] (34)   at (4,2)  {3};
    \node [vertex] (123)  at (0,3)  {};
    \node [vertex] (124)  at (2,3)  {};
    \node [vertex] (134)  at (3,3)  {};
    \node [vertex] (234)  at (4,3)  {};
    \node [vertex] (1234) at (3,4)  {};
    \draw \hasselines ;
  \end{tikzpicture}
 \begin{tikzpicture}[yscale=.6,xscale=.7]
    \node [vertex] (0)    at (0,0)  {0};
    \node [vertex] (1)    at (-1,1) {1};
    \node [vertex] (2)    at (0,1)  {1};
    \node [vertex] (3) [pin=-30:C]   at (1,1)  {1};
    \node [vertex] (4)    at (3,1)  {1};
    \node [vertex] (12)   at (-1,2) {2};
    \node [vertex] (13) [pin=120:AC]  at (0,2)  {2};
    \node [vertex] (23)   at (1,2)  {2};
    \node [vertex] (14)   at (2,2)  {2};
    \node [vertex] (24)   at (3,2)  {3};
    \node [vertex] (34)   at (4,2)  {3};
    \node [vertex] (123)  at (0,3)  {3};
    \node [vertex] (124)  at (2,3)  {4};
    \node [vertex] (134) [pin=120:ACD] at (3,3)  {4};
    \node [vertex] (234)  at (4,3)  {5};
    \node [vertex] (1234) at (3,4)  {};
    \draw \hasselines ;
  \end{tikzpicture}
  \caption{Three stages in the tabulation of $\ind(S)$ for all
    $S\subseteq N$ bottom-up. %
    For example, the value of $\ind(\{\mathsf A,\mathsf C,\mathsf
    D\})$ is given by \eqref{eq: ind(S)} with $v=\mathsf D$ as
    $\ind(\{\mathsf A,\mathsf C\}) + \ind (\{\mathsf C\}) + 1 = 4$.}
\end{figure}

\paragraph{Perspective.}

The \emph{brute force} solution for graph colouring tries all $k^n$
assignments of colours to the nodes, which is slower for $k\geq
4$. %
Another approach is \emph{dynamic programming over the subsets}
\cite{Lawler}, based on the idea that $G$ can be $k$-coloured if and
only if $G[N\setminus S]$ can be $(k-1)$-coloured for some nonempty independent set
$S$. %
That algorithm also runs within a polynomial factor of $3^n$, but uses
exponential space. %
In summary, the inclusion--exclusion approach is faster than brute
force, and uses less space than dynamic programming over the
subsets. %
The insight that this idea applies to a wide range of sequencing and
packing problems goes back to Karp \cite{Karp82}, the application to
graph colouring is from \cite{BH08}.

We use a space--time trade-off to reducing the exponential running time
factor from $3^n$ to $2^n$, %
applying dynamic programming to tabulate the decrease-and-conquer
recurrence \eqref{eq: ind(S)}, based on \cite{BHKTA}. %
Recurrence \eqref{eq: ind(S)} depends heavily on the structure of
independent sets; a more general approach is shown in \S\ref{sec:
  yates}.

The two strategies for computing $\ind(S)$ represent extreme cases of
a space--time tradeoff that can be balanced \cite{BHKK-lp}.


\section{Perfect matchings in bipartite graphs.}
\label{sec: permanent}
Consider a bipartite graph with bipartition $(N,N)$, where
$N=\{1,\ldots,n\}$, and edge set $E\subseteq N\times N$. %
A \emph{perfect matching} is an edge subset $M\subseteq E$ that
includes every node as an endpoint exactly once. %
See Fig.~\ref{fig: bipartite perfect matchings} for some
interpretations. %
\newcommand{\chessboard}[3]{
  \draw[fill=#1] (-.5,.5) rectangle +(1,1);
  \draw[fill=#2] (.5,-.5) rectangle +(1,1);
  \draw[fill=#3] (1.5,-.5) rectangle +(1,1);
  \draw[fill=#2] (.5,.5) rectangle +(1,1);
  \draw[fill=#1] (-.5,1.5) rectangle +(1,1);
  \draw[fill=#2] (.5,1.5) rectangle +(1,1);
  \draw[fill=#3] (1.5,1.5) rectangle +(1,1);
  \useasboundingbox (-.5,-.5) rectangle (2.5,2.5);
}
\newcommand{\rooksat}[3]{
  \node at (#1,0) {\brook};
  \node at (#2,1) {\brook};
  \node at (#3,2) {\brook};
}
\newcommand{\chess}[6]{
  \begin{tikzpicture}[scale=.265]
    \chessboard{#1}{#2}{#3}\rooksat{#4}{#5}{#6}
  \end{tikzpicture}
}
\tikzstyle{every loop}=[shorten >=0mm]
\tikzstyle{matchedge}=[line width=2pt]
\newcommand{\coltwo}{white}
\newcommand{\colfour}{white}
\newcommand{\colsix}{white}
\newcommand{\drawvertices}{%
  \matrix [matrix of nodes, inner sep=0pt, nodes={circle,draw,fill=white,inner sep=0pt,minimum size=3pt}, row sep=3.5pt, column sep=9pt,ampersand replacement=\&]%
  {%
    |(1)| \& |(2)|\\%
    |(3)| \& |(4)|\\%
    |(5)| \& |(6)|\\%
  };}
\newcommand{\drawedges}{
  \foreach \u/\v in {1/2, 1/4,1/6,3/2,3/4,5/6,5/4}
  \draw[lblue, semithick] (\u)--(\v);}
\newenvironment{dgraph}{
  \begin{scope}[line cap=rect]
  \drawvertices
  \drawedges}{\drawvertices
  \draw (2) node[fill=\coltwo, inner sep=0pt, minimum size=3pt,draw,
  circle] {};
  \draw (4) node[fill=\colfour, inner sep=0pt, minimum size=3pt,draw,
  circle] {};
  \draw (6) node[fill=\colsix, inner sep=0pt, minimum size=3pt,draw, circle] {};
\end{scope}}
\newenvironment{drawgraph}{\begin{tikzpicture}\begin{dgraph}
}{\end{dgraph}
\end{tikzpicture}%
\renewcommand{\coltwo}{white}%
\renewcommand{\colfour}{white}%
\renewcommand{\colsix}{white}%
}

\begin{SCfigure}
\tikzstyle{blacknumbers}=[black,font=\sffamily\bfseries]
$\begin{array}{cccc}
\begin{drawgraph}\end{drawgraph}
&
\begin{drawgraph}
  \foreach \u/\v in {1/2, 3/4,5/6} \draw[matchedge] (\u)--(\v);
\end{drawgraph}
&
\begin{drawgraph}
  \foreach \u/\v in {1/6, 3/2,5/4} \draw[matchedge] (\u)--(\v);
\end{drawgraph}
&
\begin{drawgraph}
  \foreach \u/\v in {1/4, 3/2,5/6} \draw[matchedge] (\u)--(\v);
\end{drawgraph}
\\[1ex]
\begin{tikzpicture}
  \matrix[matrix of nodes,inner sep=2pt, outer sep= 0pt,font=\small\sf]{
    1& 1& 1\\
    1& 1& 0\\
    0& 1& 1\\
  };
\end{tikzpicture}
&
\begin{tikzpicture}
  \matrix[matrix of nodes,inner sep=2pt,font=\small\sf, color=lblue]{
    |[blacknumbers]|1& 1& 1\\
    1& |[blacknumbers]|1& 0\\
    0& 1& |[blacknumbers]|1\\
  };
\end{tikzpicture}
&
\begin{tikzpicture}
  \matrix[matrix of nodes,inner sep=2pt,font=\small\sf, color=lblue]{
    1& 1& |[blacknumbers]| 1\\
    |[blacknumbers]|1& 1& 0\\
    0& |[blacknumbers]|1& 1\\
  };
\end{tikzpicture}
&
\begin{tikzpicture}
  \matrix[matrix of nodes,inner sep=2pt,font=\small\sf, color=lblue]{
    1& |[blacknumbers]|1& 1\\
    |[blacknumbers]|1& 1& 0\\
    0& 1& |[blacknumbers]| 1\\
  };
\end{tikzpicture}
\\[1ex]
\begin{tikzpicture}[inner sep=1pt,every loop=[]]
\node (1) [circle, draw] at (90:.33) {\scriptsize\sf 1};
\node (2) [circle, draw] at (330:.33) {\sf\scriptsize 2};
\node (3) [circle, draw] at (210:.33) {\sf\scriptsize 3};
\path[->] (1) edge[out=60,in=120,loop, looseness=7] (1)  edge [bend left] (2) edge (3)
          (2) edge [out=360,in=300,loop,looseness=7 ] (2) edge [bend left] (1)
          (3) edge [out=180,in=240,loop,looseness=7 ] (3) edge (2);
\end{tikzpicture}
&
\begin{tikzpicture}[inner sep=1pt,every loop=[]]
\node (1) [circle, draw] at (90:.33) {\scriptsize\sf 1};
\node (2) [circle, draw] at (330:.33) {\scriptsize\sf 2};
\node (3) [circle, draw] at (210:.33) {\scriptsize\sf 3};
\path[->] (1) edge[out=60,in=120,loop,very thick, looseness=7] (1)
edge [lblue,bend left] (2) edge [lblue] (3)
          (2) edge [out=360,in=300,very thick,loop,looseness=7 ] (2)
          edge [bend left, lblue] (1)
          (3) edge [out=180,in=240,very thick,loop,looseness=7 ] (3) edge [lblue](2);
\end{tikzpicture}
&
\begin{tikzpicture}[inner sep=1pt,every loop=[]]
\node (1) [circle, draw] at (90:.33) {\scriptsize\sf 1};
\node (2) [circle, draw] at (330:.33) {\scriptsize\sf 2};
\node (3) [circle, draw] at (210:.33) {\scriptsize\sf 3};
\path[->] (1) edge[out=60,in=120,loop, looseness=7,lblue] (1)  edge [bend
left,lblue] (2) edge [very thick] (3)
          (2) edge [out=360,in=300,loop,looseness=7 ,lblue] (2) edge [bend
          left, very thick] (1)
          (3) edge [out=180,in=240,loop,looseness=7, lblue] (3) edge [very
          thick] (2);
\end{tikzpicture}
&
\begin{tikzpicture}[inner sep=1pt,every loop=[]]
\node (1) [circle, draw] at (90:.33) {\scriptsize\sf 1};
\node (2) [circle, draw] at (330:.33) {\scriptsize\sf 2};
\node (3) [circle, draw] at (210:.33) {\scriptsize\sf 3};
\path[->] (1) edge[out=60,in=120,loop, looseness=7,lblue] (1)  edge [bend
left, very thick] (2) edge [lblue] (3)
          (2) edge [out=360,in=300,loop,looseness=7 ,lblue] (2) edge [bend
          left, very thick] (1)
          (3) edge [out=180,in=240,loop,looseness=7, very thick ] (3) edge [lblue] (2);
\end{tikzpicture}
\\[1ex]
\begin{tikzpicture}[scale=.265]
  \chessboard{\lblue}{\lblue}{\lblue}
\end{tikzpicture}
&
\chess{\lblue}{\lblue}{\lblue}{2}{1}{0}
&
\chess{\lblue}{\lblue}{\lblue}{1}{0}{2}
&
\chess{\lblue}{\lblue}{\lblue}{2}{0}{1}
\end{array}$
\caption{\label{fig: bipartite perfect matchings}%
  Row 1: %
  A bipartite graph and its three perfect matchings. %
  \enspace Row~2: %
  In the graph's adjacency matrix $A$, every perfect matching
  corresponds to a permutation $\pi$ for which $A_{i,\pi(i)}=1$ for
  all $i\in [n]$. %
  \enspace Row~3: %
  In the directed $n$-node graph defined by $A$, every perfect
  matching corresponds to a directed cycle partition.  %
  \enspace Bottom row: %
  an equivalent formulation in terms of non-attacking rooks on a
  chess board with forbidden positions.}
\end{SCfigure}
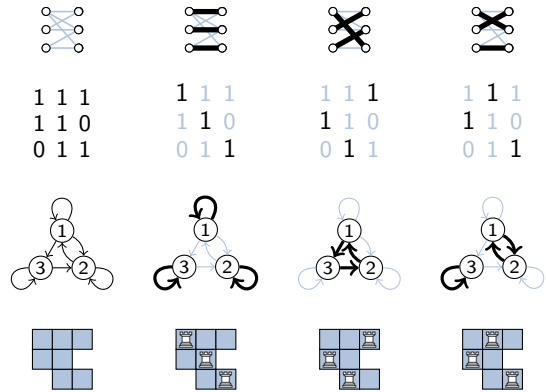

\begin{figure}[tb]
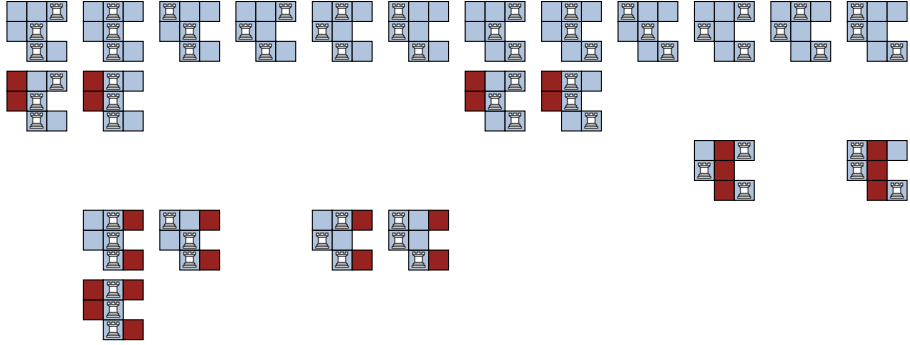

\begin{tabular}{ccccccccccccc}
\chess{\lblue}{\lblue}{\lblue}{1}{1}{2} &
\chess{\lblue}{\lblue}{\lblue}{1}{1}{1} &
\chess{\lblue}{\lblue}{\lblue}{1}{1}{0} &
\chess{\lblue}{\lblue}{\lblue}{1}{0}{2} &
\chess{\lblue}{\lblue}{\lblue}{1}{0}{1} &
\chess{\lblue}{\lblue}{\lblue}{1}{0}{0} &
\chess{\lblue}{\lblue}{\lblue}{2}{1}{2} &
\chess{\lblue}{\lblue}{\lblue}{2}{1}{1} &
\chess{\lblue}{\lblue}{\lblue}{2}{1}{0} &
\chess{\lblue}{\lblue}{\lblue}{2}{0}{2} &
\chess{\lblue}{\lblue}{\lblue}{2}{0}{1} &
\chess{\lblue}{\lblue}{\lblue}{2}{0}{0}
\\
\chess{rred}{\lblue}{\lblue}{1}{1}{2} &
\chess{rred}{\lblue}{\lblue}{1}{1}{1} &
&&&&
\chess{rred}{\lblue}{\lblue}{2}{1}{2} &
\chess{rred}{\lblue}{\lblue}{2}{1}{1} &
&&&&
\\
&&&&&&&&&
\chess{\lblue}{rred}{\lblue}{2}{0}{2} 
&&
\chess{\lblue}{rred}{\lblue}{2}{0}{0}
\\
&
\chess{\lblue}{\lblue}{rred}{1}{1}{1} &
\chess{\lblue}{\lblue}{rred}{1}{1}{0} &
&
\chess{\lblue}{\lblue}{rred}{1}{0}{1} &
\chess{\lblue}{\lblue}{rred}{1}{0}{0} &
&&&&&\\
&
\chess{rred}{\lblue}{rred}{1}{1}{1} &
\end{tabular}
\caption{\label{fig: Ryser}\binoppenalty = 0 \relpenalty=0
  Inclusion--exclusion for non-attacking rooks. %
  The top row shows all $12=3\cdot 2\cdot 2$ ways to place exactly one
  rook in every board line. %
  Every row shows the possible placements in the vertical lines given
  by $S\subseteq \{1,2,3\}$. %
  We omit the rows whose contribution vanishes, namely $S=\{1\}$,
  $S=\{3\}$ and $S=\emptyset$. %
  Of particular interest is the second column, which is subtracted
  twice and later added again. %
  The entire calculation is $12-4-2-4+1+0+0-0=3$.}

\end{figure}

The \emph{Ryser formula} for counting the perfect matchings in such a
graph can be given as
\begin{equation}\label{eq: ryser bip} 
  \sum_{\pi\in S_n}\prod_{i=1}^{n} [i\pi(i) \in E] =
  \sum_{S\subseteq N} (-1)^{|N\setminus S|} \prod_{i=1}^n \sum_{j\in S}
  [ij\in E] \,,
\end{equation}
where $S_n$ denotes the set of permutations from $N$ to $N$. %
The left hand side succinctly describes the problem as iterating over
all permutations and checking if the corresponding edges (namely,
$1\pi(1)$, $2\pi(2)$, $\ldots$, $n\pi(n)$) are all in $E$. %
Direct evaluation would require $n!$ iterations. %
The right hand side provides an equivalent expression that can be
evaluated in time $O(2^nn^2)$, see Fig.~\ref{fig: Ryser}.

\begin{pf}[Proof of \eqref{eq: ryser bip}]
  For fixed $i\in N$, the value $\sum_{j\in S}[ij\in E]$ counts the
  number of $i$'s neighbours in $S\subseteq N$. %
  Thus the expression \begin{equation}\label{eq: ryser contrib}
   \prod_{i=1}^n\sum_{j\in S}[ij\in E]
   \end{equation}
   is the number of ways every node $i\in N$ can choose a neighbour
   from $S$. %
   (This allows some nodes to select the same neighbour.) %
   Consider such a choice as a mapping $g\colon N\rightarrow N$, not
   necessarily onto, with image $R=g(N)$. %
   The contribution of $g$ to \eqref{eq: ryser contrib} is 1 for every
   $S\supseteq R$, and its total contribution to the right hand
   side of \eqref{eq: ryser bip} is, using \eqref{eq: pie},
  \[
  \sum_{R\subseteq S\subseteq N} (-1)^{|N\setminus S|} \cdot 1 =
  [g(N)=N]\,.
  \]
  Thus $g$ contributes if and only if it is a permutation.
\end{pf}

\paragraph{Perspective.}

Bipartite matching is an example of a sequencing problem, where
inclusion--exclusion replaces an enumeration over permutations,
$\sum_{\pi\in S_n}$ by an alternating enumeration over
subsets $\sum_{S\subseteq N}(-1)^{|N\setminus S|}$ of functions with
restricted range. %
Typically, this reduces a factor $n!$ in the running time to $2^n$. %
One can express the idea algebraically like this: 
\begin{equation}\label{eq: ie for functions}
\begin{split}
\sum_{\substack{f\colon N\rightarrow N\\ f(N)=N}} [\,\cdots] & =
  \sum_R [R=N] \sum_{\substack{f\colon N\rightarrow N\\ f(N)=R}}
    [\,\cdots] \\
    & =
    \sum_R \sum_S [R\subseteq S] (-1)^{|N\setminus S|}
    \sum_{\substack{f\colon N\rightarrow N\\ f(N)=R}} [\,\cdots]\\
   &=
   \sum_S (-1)^{|N\setminus S|} \sum_R [R\subseteq S]
   \sum_{\substack{f\colon N\rightarrow N\\ f(N)=R}} [\,\cdots]\\
     &=
     \sum_S (-1)^{|N\setminus S|}
     \sum_{f\colon N\rightarrow S} [\,\cdots] \,.
\end{split}
\end{equation}

Ryser's formula is normally given in a more general form, for the
\emph{permanent} $\sum_\pi \prod_i A_{i\pi(i)}$ of a matrix, where the
entries can be other than just 0 and 1. %
The running time can be improved to $O(2^nn)$ arithmetic operations by
iterating over $N$ in Gray code order. %

Ryser's formula \cite{Ryser:1963} is a very well-known result in
combinatorics and appears in many textbooks. %
However, it is easy to achieve running time $O(2^nn)$ using dynamic
programming over the subsets, at the expense of space $O(2^n)$. %
This is the standard approach to sequencing problems
\cite{Bell62,HeKa62}, and appears as an exercise in Knuth
\cite[pp. 515-516]{Knuth:1998}, but usually not in the combinatorics
literature. %
We will witness the opposite methodological preferences in \S\ref{sec:
  HC}. %
Inclusion--exclusion-based algorithms for the permanent of non-square
matrices in semirings are described in \cite{BHKK-perm}.

\section{Perfect matchings in general graphs.}
\label{sec: general graphs}
We turn to graphs that are not necessarily bipartite.  %
In general, the number of perfect matchings in a graph with an even
number $n$ of nodes $N$ is
\begin{equation}\label{eq: pm general}
\sum_{S\subseteq N} (-1)^{|N\setminus S|} \binom{e[S]}{n/2},
\end{equation}
where $e[S]$ denotes the number of edges between nodes in
$S\subseteq N$.

\begin{sidewaysfigure}
\newenvironment{graph}{
\begin{tikzpicture}[scale=.19, inner sep = 1.2pt]
  \node (0) [circle,draw] at (0,2) {};
  \node (1) [circle,draw] at (0,1) {};
  \node (2) [circle,draw] at (0,0) {};
  \node (3) [circle,draw] at (1,2) {};
  \node (4) [circle,draw] at (1,1) {};
  \node (5) [circle,draw] at (1,0) {};
  \begin{scope}
  \path (1) edge (0) 
            edge (2) 
            edge (4) 
            edge (5)
        (3) edge (0) 
            edge (4)
        (2) edge (5);
  \end{scope}
\begin{scope}[ultra thick, line cap=round]
}{\end{scope}\end{tikzpicture}}
\[
\begin{array}{@{\extracolsep{3pt}}cccccccccccccccccccccccccccccccccccc}
\input{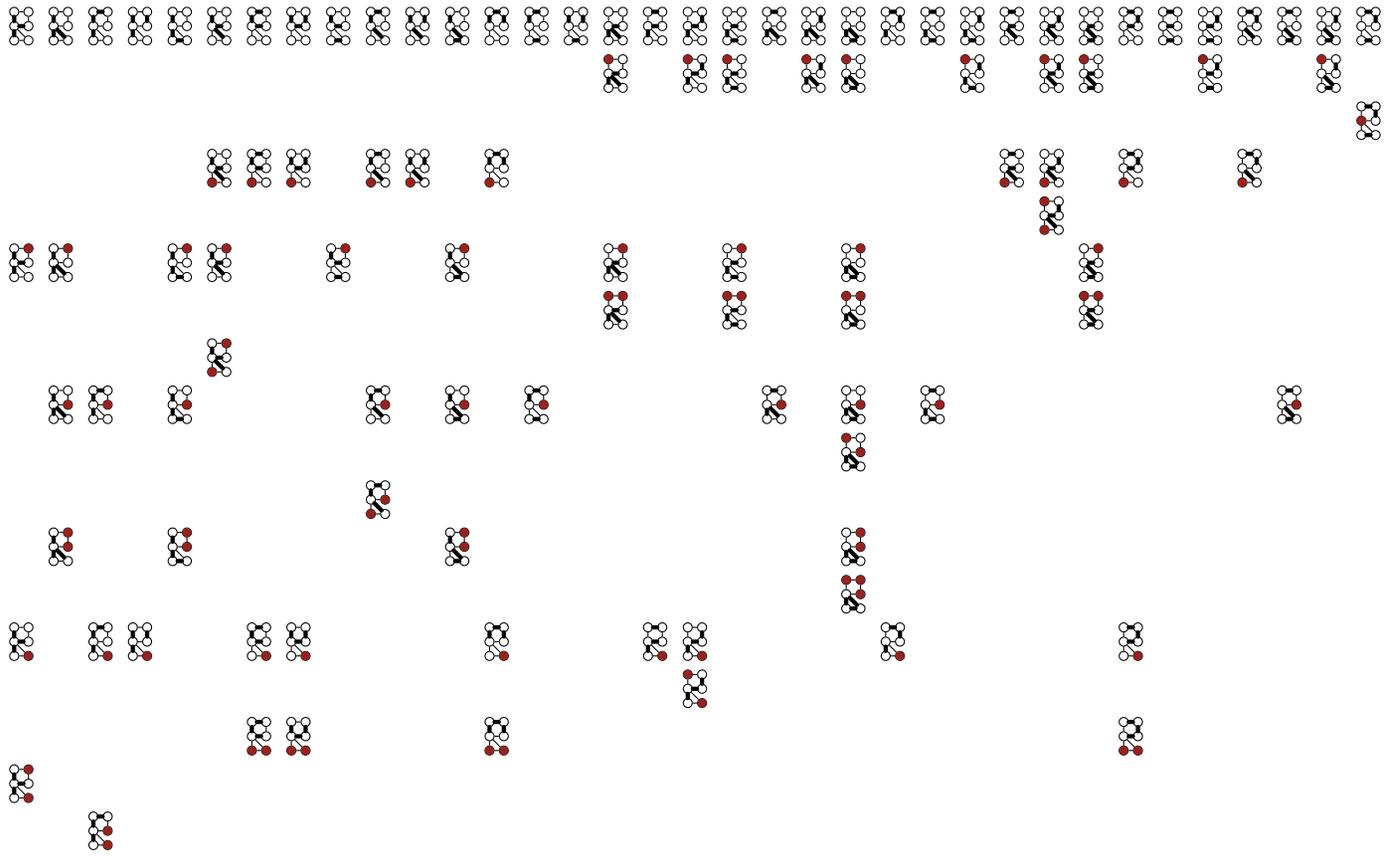}
\end{array}\]
\caption{\label{fig: pm}The perfect matching algorithm for a graph
  with $n=6$ and $m=7$. %
  There are $\binom{7}{3}=35$ ways to pick 3 edges out of 7, shown in
  the top row. %
  The triangle appears in 7 other terms (4 negative, 3 positive), the
  two perfect matchings appear only once.  }
\end{sidewaysfigure}

\begin{pf}
  The term $\binom{e[S]}{n/2}$ counts the number of ways to select $n/2$
  distinct edges with endpoints in $S$. %
  (The edges are distinct, but may share nodes.) %
  Consider such a selection $F\subseteq E$ and let $R=\bigcup_{uv\in F}
  \{u,v\}$ denote the nodes covered by the selected edges. %
  The total contribution of $F$ to the right hand side of \eqref{eq:
    pm general} is
  \[ \sum_{R\subseteq S\subseteq N} (-1)^{|N\setminus S|}
  \binom{e[S]}{n/2} = [R=N]\,,\] using \eqref{eq: pie}. %
  Thus, $F$ contributes 1 if and only if it covers all nodes.  %
  Since $F$ contains $n/2$ edges, $F$ it must be a perfect
  matching.  %
\end{pf}

The running time is within a polynomial factor of $2^n$, and the space
is polynomial; see Fig.~\ref{fig: pm}. %

\paragraph{Perspective.}

Perfect matchings in general graphs is a packing or partitioning
problem, %
while the bipartite case was a a sequencing problem %
and the graph colouring example in \S\ref{sec: colouring} was a
covering problem. %
(Admittedly, the distinction between these things is not very
clear.) %
The application is form \cite{BH08}, which also contains another
space--time trade-off based on matrix multiplication.

The point of the large in example in Fig.~\ref{fig: pm} is to
illustrate the intuition that inclusion--exclusion is a
\emph{sieve}. %
We start with a large collection of easy-to-compute objects (the top
row in Fig.~\ref{fig: pm}), and let the alternating sum perform a
cancellation that sifts through the objects and keeps only the
interesting ones in the sieve.

\section{Inclusion--exclusion for sets.}
\label{sec: ie sets}
\begin{wrapfigure}[4]{r}{1.5cm}
\begin{tikzpicture}[scale=.5]
  \begin{scope}
    \clip (0,0) circle (1);
    \fill [lblue] (1,0) circle (1);
  \end{scope}
  \draw (0,0) circle (1);
  \draw (1,0) circle (1);
\end{tikzpicture}
\end{wrapfigure}
If two sets $A$ and $B$ have no elements in common, then we have the
principle of \emph{addition}: \(|A\cup B|=|A|+|B|\,\). %
In general, the equality does not hold, and all we have is \(|A\cup
B|\leq |A|+|B|\,\). %
Observing that every element of $A\cap B$ is counted exactly twice on
the right hand side allows us subtract the error term:\[|A\cup B|=
|A|+|B|-|A\cap B|\,,
\] often called the \emph{principle of
  inclusion--exclusion}. 

\begin{wrapfigure}[7]{r}{2cm}
\vspace{-3ex}
\begin{tikzpicture}[scale=.7]
  \begin{scope}
    \clip (90:.5) circle (1);
    \fill [lblue] (210:.5) circle (1);
  \end{scope}
  \begin{scope}
    \clip (210:.5) circle (1);
    \fill [lblue] (330:.5) circle (1);
  \end{scope}
  \begin{scope}
    \clip (330:.5) circle (1);
    \fill [lblue] (90:.5) circle (1);
  \end{scope}
  \begin{scope}
    \clip (90:.5) circle (1);
    \clip (210:.5) circle (1);
    \fill [dblue] (330:.5) circle (1);
  \end{scope}
  \draw (90:.5) circle (1);
  \draw (210:.5) circle (1);
  \draw (330:.5) circle (1);
\end{tikzpicture}
\end{wrapfigure}

Actually, that's just a special case, the formula is elevated to a
principle by generalising to more sets. %
For three sets, the formula \[|A\cup B\cup C|= |A|+|B|+|C|-|A\cap
B|-|A\cap C|-|B\cap C|+|A\cap B\cap C|\] can be verified by staring at
a Venn diagram. %
The right-hand side contains all the possible intersections of $A$,
$B$, and $C$, with signs depending on how many sets intersect. %
Generalising this leads us to
\begin{equation}\label{eq: ie sets uncompl}
  |A_1\cup\cdots\cup A_n| = \sum_{\emptyset\neq S\subseteq N}
  (-1)^{|S|+1} \bigl|\bigcap_{i\in S}  A_i\bigr|,
\end{equation}
where $N=\{1,\ldots,n\}$. %
Equivalently, the number of elements not in any $A_i$ is
\begin{equation}\label{eq: ie sets}
  \bigl|\overline{A_1\cup\cdots\cup A_n}\bigr| = \sum_{S\subseteq N}
  (-1)^{|S|} \bigl|\bigcap_{i\in S}  A_i\bigr|,
\end{equation}
with the usual convention that the `empty' intersection $\bigcap_{i\in
  \emptyset} A_i$ equals the universe from which the sets are taken.

\begin{pf}[Proof of \eqref{eq: ie sets}]
  We consider the contribution of every element $a$. %

  Let $T=\{\, i\in N\colon a\in A_i\,\}$ denote the index set of the
  sets containing $a$. %
  The contribution of $a$ to the left hand side of \eqref{eq: ie sets}
  is $[T=\emptyset]$. %
  To determine its contribution to the right hand side, we observe
  that $a$ belongs to the intersection $\bigcap_{i\in T} A_i$ and all
  its sub-intersections, %
  so it contributes 1 to all corresponding terms. %
  More precisely, the total contribution of $a$ is given by
  \[ \sum_{S\subseteq T} (-1)^{|S|}= (-1)^{|T|}  \sum_{S\subseteq T}
  (-1)^{|T\setminus S|} 
    = (-1)^{|T|} [T=\emptyset]=  [T=\emptyset]\,,\]
    using \eqref{eq: pie} with $R=\emptyset$.
\end{pf}

\paragraph{Perspective.}
Expressions \eqref{eq: ie sets uncompl} and \eqref{eq: ie sets} are
the standard textbook presentation of inclusion--exclusion. %
We derived them from \eqref{eq: pie} with $R=\emptyset$. %
Let us show the opposite derivation to see that the two formulations
are equivalent. %

Let $T$ be a nonempty, finite set and write $T=\{1,\ldots, n\}$. %
Consider the family of identical subsets $A_i=\{1\}$ for all $i\in
T$. %
Their union and every nonempty intersection is $\{1\}$.  %
Thus, from \eqref{eq: ie sets uncompl}, %
\[ 1= \sum_{\emptyset \neq S\subseteq T} (-1)^{|S|+1}\cdot
1\,,\] which gives \eqref{eq: pie} for $R=\emptyset$ after
subtracting 1 from both sides.

\section{Hamiltonian paths.}
\label{sec: HC}

A \emph{walk} in a graph is a sequence of neighbouring nodes
$v_1,\ldots, v_k$. %
Such a walk is a \emph{path} if every node appears at most once, and a
path is \emph{Hamiltonian} if it includes every vertex in $G$. %
For ease of notation we also assume that all Hamiltonian paths start
in node $v_1=1$.

Given a graph $G$ on $n$ nodes $N$ let $a(X)$ denote the number of
walks of length $n$ that start in $1$ and `avoid' the nodes in 
$X\subseteq V$, i.e., walks of the form $1=v_1,\ldots, v_n$ with
$v_i\notin X$ for all $1\leq i\leq n$. %
Then the number of Hamiltonian paths in $G$ is $a(\emptyset)$. %

Let $A_i$ denote the walks that avoid $\{i\}$. %
Then $a(\emptyset)=|\bigcup_{i\in N} A_i|$ and $a(X)=|\bigcap_{i\in X}
A_i|$. %
Thus, from \eqref{eq: ie sets}, we have
\[a(\emptyset)=\sum_{X\subseteq N} (-1)^{|X|} a(X)\,.\]

For every $X$, the value $a(X)$ can be computed in polynomial time
using dynamic programming (over the lengths and endpoints, not over
the subsets). %
For $t\in V$ and $k=1,\ldots,n$ let for a moment $a^k(X,t)$ denote the
number of walks of the form $1=v_1,\ldots,v_k=t$ with $v_i\notin X$. %
Then we can set $a^1(X,v)=[v=1]$ and
\[ a^{k+1}(X,t)= \sum_{v\in V} a^k(X,v)[vt\in E]\,.\] %
The total time to compute $a(X)=\sum_{t\in V} a(X,t)$ becomes
$O(n^2|E|)$, using polynomial space.  %
It follows that Hamiltonicity in an $n$-node graph can be decided
(in fact, counted) in time $O(2^nn^2|E|)$ and linear space.

\paragraph{Perspective.}

Hamiltonicity is one of the earliest explicitly algorithmic
applications of inclusion--exclusion. %
It appears in \cite{Karp82}, but implicitly already in \cite{KoGK77},
where it is described for the traveling salesman problem with bounded
integer weights. %
Both these papers have lived in relative obscurity until recently, for
example the TSP result has been both reproved and called `open' in a
number of places.

Hamiltonicity is also the canonical application of another algorithmic
technique, \emph{dynamic programming over the subsets}
\cite{Bell62,HeKa62}, which yields an algorithm with similar time
bounds but exponential space. %
Thus, we can observe a curious cultural difference in the default
approach to hard sequencing problems: %
Dynamic programming is the well-known solution to Hamiltonicity, while
the inclusion--exclusion formulation is often overlooked. %
For the permanent (\S\ref{sec: permanent}), the situation is reversed.



\section{Steiner tree.}

For a graph $G=(N,E)$ and a subset $\{t_1,\ldots, t_k\}\subseteq N$ of
nodes called \emph{terminals}, a \emph{Steiner tree} is a tree in $G$
that contains all terminals. %
We want to determine the smallest size of such a tree.

We consider a related structure that is to a tree what a walk is to a
path. %
A \emph{willow} $W$ consists of a multiset of nodes $S(W)$ from $N$
and a parent function $p\colon S(W)\rightarrow S(W)$, such that
repeated applications of $p$ end in a specified \emph{root} node $r\in
S(W)$. %
The size of $W$ is the number of nodes in $S(W)$, counted with
repetition. %

Every tree can be turned into a willow, by selecting an arbitrary root
and orienting the edges towards it, but the converse is not true. %
However, a minimum size willow over a set of nodes is a tree: %
Assume a node appears twice in $W$. %
Remove one of its occurrences $u\in S(W)$, not the root node, and
change the parent $p(v)$ of all $v$ with $p(v)=u$ to $p(v)=p(u)$. %
The resulting willow is smaller than $W$ but spans the same nodes.
Finally, when all repeated nodes are removed, $p$ defines a tree.

Thus, it suffices to look for a size-$l$ willow $W$ that includes all
terminals, for increasing $l = k,\ldots, n$. %
Set $A_{i}$ to be the set of willows of size $l$ that avoid $t_i$. %
Then, from \eqref{eq: ie sets}, the number of willows of size $l$
that include all terminals is
 \[ \sum_{X\subseteq K} (-1)^{|X|} a^l(X)\,,
\]
where $a^l(X)=|\bigcap_{i\in X} A_i|$ is the number of willows of size
$l$ that avoid the terminals in $X$.

Again, we can use dynamic programming to compute $a^l(X)$ for given
$X$. %
For all $X\subseteq V$ and $u\notin X$ let $a^l(X,u)$ denote the
number willows of size $l$ that avoid $X$ and whose root is $u\notin
X$. %
Then $a^1(X,u)= 1$ and
\[ a^k (X,u)=\sum_{uv\in E}\sum_{i=1}^{k-1} a^i(X,u)a^{k-i}(X,v)\,.\]

\paragraph{Perspective.} 
This application is from \cite{Nede09}. %
The role of inclusion--exclusion is slightly different from the
Hamiltonian path construction in the previous subsection, because
we have no control over the size of the objects we are sieving for. %

There, we sifted through all walks of length $n$. %
What was left in the sieve were the walks of length $n$ that visit all
$n$ nodes. %
Thus, every node appears exactly once, so that sieve contained exactly
the desired solutions, i.e., the Hamiltonian paths.

Here, we sift through all willows of size $l$. %
What is left in the sieve are the willows that visit all terminals. %
For given $l$, these are not necessarily trees. %
Instead, correctness hinges on the fact that we already sifted through
willows of smaller size. %
To strain the metaphor, we use increasingly fine sieves until we find
something.

\section{Long paths.}

Consider a graph $G=(N,E)$ and integer  $k\leq n$. %
We want to detect if $G$ has a path of length $k$. %
Inspired by the Hamiltonicity construction in \S\ref{sec: HC} we look
at all walks $(w_1,\ldots,w_k)$ on $k$ nodes. %
For expository reasons we again stipulate that all walks begin in a
fixed node $w_1=1$. %
Write $K=\{1,\ldots,k\}$.

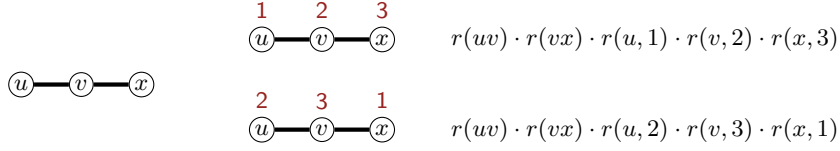
\begin{figure}[tb]
\[
\begin{tikzpicture}[scale=.8]
   \begin{scope}
      \node[circle, draw, inner sep=1] (1) at (0,0) {$u$};
      \node[circle, draw, inner sep=1] (2) at (1,0) {$v$};    
      \node[circle, draw, inner sep=1] (3) at (2,0) {$x$};
      \draw [ultra thick] (1)-- (2)-- (3);
    \end{scope}
   \begin{scope}[xshift=4cm,yshift=.75cm]
      \node[circle, draw, inner sep=1] (1) at (0,0)[label=above:1] {$u$};
      \node[circle, draw, inner sep=1] (2) at (1,0)[label=above:2] {$v$};    
      \node[circle, draw, inner sep=1] (3) at (2,0)[label=above:3] {$x$};
      \draw [ultra thick] (1)-- (2)-- (3);
      \node[anchor=west] at(3,0) {\small $r(uv)\cdot r(vx)\cdot r(u,1)\cdot r(v,2) \cdot r(x,3)$};
    \end{scope}
   \begin{scope}[xshift=4cm, yshift=-.75cm]
      \node[circle, draw, inner sep=1] (1) at (0,0)[label=above:2] {$u$};
      \node[circle, draw, inner sep=1] (2) at (1,0)[label=above:3] {$v$};    
      \node[circle, draw, inner sep=1] (3) at (2,0)[label=above:1] {$x$};
      \draw [ultra thick] (1)-- (2)-- (3);
      \node[anchor=west] at(3,0){\small$r(uv)\cdot r(vx)\cdot r(u,2)\cdot r(v,3) \cdot r(x,1)$};
    \end{scope}
\end{tikzpicture}
\]
\caption{\label{fig: labelled walks}Left: The path $W=(u,v,x)$. Middle: The nodes of $W$ labelled
  with two permutations. Right: The terms associated with $W$ and the
  two permutations.}
\end{figure}

For every edge $e$ pick a random value $r(e)$. %
For every vertex $v$ and integer $k\in K$ pick a random value
$r(v,k)$. %
With foresight, the values are chosen uniformly at random form a
finite field $F$ of characteristic $2$ and size at least $2k(k-1)$. %
All computation is in this field. %
For every walk $W=(w_1,\ldots, w_k)$ starting in $w_1=1$ and every
function $\phi\colon K\rightarrow K$ define the term
\begin{equation}
  \label{eq: monomial}
  p(W,\phi) = \biggl(\,\prod_{i=1}^{k-1}
  r(w_iw_{i+1})\biggr)
  \biggl(\,\prod_{i=1}^k  r(w_i,\phi(i))
  \biggr)\,,
\end{equation}
see Fig.~\ref{fig: labelled walks}.

Consider the sum over all walks $W$ in $G$ and all permutations
$\pi\in S_k$,
\begin{equation}
  \label{eq: p}
p(G)=  \sum_{\pi\in S_k}\sum_W p(W,\pi)\,.
\end{equation}
We will show below that 
\begin{equation}\label{eq: correctness}
  \Pr (\, p(G)=0 \,) \,
  \begin{cases}
    < \frac{1}{2}\,, & \text{if $G$ contains a $k$-path}\,;\\
    = 0\,, & \text{otherwise}\,.
\end{cases}
\end{equation}
where the probability is taken over the random choices of $r$.

To compute \eqref{eq: p}, we first recognise a summation over a
permutation and replace it by an alternating sum over functions with
restricted range, as in
\eqref{eq: ie for functions}:
\begin{equation*}
  \sum_{\pi\in S_k}\sum_W p(W,\pi) =
  \sum_{S\subseteq K} (-1)^{|K\setminus S|} 
  \sum_{\phi\colon K\rightarrow S}\sum_W p(W,\phi)\,.
\end{equation*}
For each $S\subseteq K$, the value of the two inner sums can be
computed efficiently using dynamic programming; we omit the details. %
The total running time is within a polynomial (in $n$) factor of
$2^k$.

\begin{pf}[Proof of \eqref{eq: correctness}]
  Consider the contribution of every walk $W$ to \eqref{eq: p}. %
  
  First assume that $W$ is non-simple and let $\pi$ be a permutation. %
  We will construct anther permutation $\rho$ such that $\pi\neq \rho$
  but $p(W,\pi)= -p(W,\rho)$. %
  Thus, summing over all permutations, the contribution of $W$
  is even and therefore vanishes in $F$. %
  To construct $\rho$, let $(i,j)$ be the first self-intersection on
  $W$, i.e., the lexicographically minimal pair with $w_i=w_j$ and
  $i<j$. %
  Set $\rho$ equal to $\pi$ except for $\rho(i)=\pi(j)$ and
  $\rho(j)=\pi(i)$. %
  
  Now assume that $W$ is a path. %
  It is useful to view \eqref{eq: p} as a polynomial in variables $
  x(e), x(v,k)$, evaluated at random points $x(e)=r(e)$,
  $x(v,k)=r(v,k)$. %
  For every permutation $\pi$, the monomial 
  \[
  \biggl(\prod_{i=1}^{k-1} x(w_iw_{i+1}) \biggr)
  \biggl(\prod_{i=1}^{k-1} x(w_i,\pi(i)) \biggr)
  \]
  is unique. %
  To see this, both $W$ and $\pi$ can be recovered from $p(W,\pi)$ by
  first reconstructing the nodes $w_1,\ldots, w_k$ in order, starting
  at $w_1=1$ and following the unique incident edge described by the
  terms $x(e)$, and then reconstructing $\pi$ from the terms
  $x(w_i,\pi(i))$. %
  Thus, \eqref{eq: p} can be viewed as a nonzero polynomial of degree
  $k(k-1)$ evaluated at $m+nk$ random points from $F$. %
  By the DeMillo--Lipton--Schwarz--Zippel lemma \cite{DL,Schw80}, it
  evaluates to zero with probability less than $k(k-1)/|F| \leq
  \frac{1}{2}$.
\end{pf}

\paragraph{Perspective.}  The construction is implicit in
\cite{BHKK-narrow} and not optimal. %
Again, the starting point is the same as for Hamiltonicity in
\S\ref{sec: HC}: %
to sieve for paths among walks, whose contribution is computed by
dynamic programming. %

However, instead of counting the number of walks, we define an
associated family of algebraic objects (namely, a multinomial defined
by the walk and a permutation) and work with these objects instead. %
Strictly speaking, we did associate algebraic objects to walks even
before, but the object was somewhat innocent: the integer 1.

There are two filtering processes at work: %
The sifting for paths among walks is performed by the the cancellation
of non-simple, permutation-labelled walks in characteristic 2,
rather than the by inclusion--exclusion sieve. %
At the danger of overtaxing the sieving metaphor, the
permutation-labelling plays the role of \emph{mercury} in gold
mining; %
inclusion--exclusion ensures that the `mercury' can be added and
removed in time $2^k$ instead of the straightforward $k!$.


\section{Yates's algorithm.}
\label{sec: yates}

Let $f\colon 2^N\rightarrow \{0,1\}$ be the the indicator function of
the nonempty independent sets in a graph. %
We will revisit the task of \S\ref{sec: compute g}, computing 
\begin{equation}\label{eq: ind as zeta}
\ind (S) = \sum_{R\subseteq S} f(R)\,,
\end{equation}
for all $S\subseteq N$.

The computation proceeds in rounds $i=1,\ldots, n$. %
Initially, set $g_0(S)= f(S)$ for all $S\subseteq N$. %
Then we have, for  $i=1,\ldots,n$,
\begin{equation}\label{eq: yates}
   g_i(S)= g_{i-1}(S)+ [i\in
  S]\cdot g_{i-1}(S\setminus\{i\})  \qquad (S\subseteq N)\,.
\end{equation}
Finally, $\ind(S)=g_n(S)$.

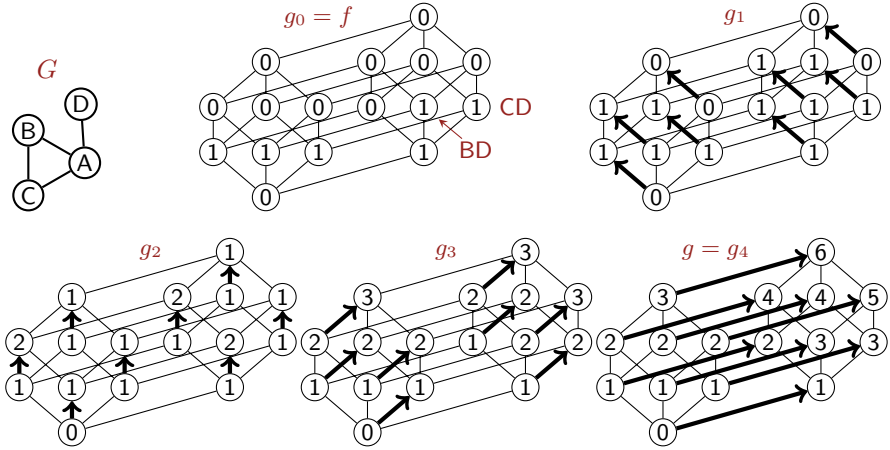
\begin{figure}
\[
  \begin{tikzpicture}[scale=.5, every node/.style={circle, fill=white,draw,thick,inner sep=1pt,font=\sf\small}]
      \node (1) at (0:1) {A};
      \node (2) at (120:1) {B};    
      \node (3) at (240:1) {C};
      \node (4) at (60:1.8) {D};
      \draw [thick](4)--(1)--(2)--(3)--(1);
      \node [rred,draw=none,fill=none,font=\normalsize] (legend) at (90:2.5) {$G$};
    \end{tikzpicture}
    \qquad\qquad
     \begin{tikzpicture}[yscale=.6,xscale=.7]
    \node [vertex] (0)    at (0,0)  {0};
    \node [vertex] (1)    at (-1,1) {1};
    \node [vertex] (2)    at (0,1)  {1};
    \node [vertex] (3)    at (1,1)  {1};
    \node [vertex] (4)    at (3,1)  {1};
    \node [vertex] (12)   at (-1,2) {0};
    \node [vertex] (13)   at (0,2)  {0};
    \node [vertex] (23)   at (1,2)  {0};
    \node [vertex] (14)   at (2,2)  {0};
    \node [vertex,pin=-45:BD] (24)   at (3,2)  {1};
    \node [vertex,label=right:CD] (34)   at (4,2)  {1};
    \node [vertex] (123)  at (0,3)  {0};
    \node [vertex] (124)  at (2,3)  {0};
    \node [vertex] (134)  at (3,3)  {0};
    \node [vertex] (234)  at (4,3)  {0};
    \node [vertex] (1234) at (3,4)  {0};
    \node [rred, overlay] (legend) at (1,4) {$g_0=f$};
    \draw \hasselines ;
  \end{tikzpicture}
\qquad
  \begin{tikzpicture}[yscale=.6,xscale=.7]
    \node [vertex] (0)    at (0,0)  {0};
    \node [vertex] (1)    at (-1,1) {1};
    \node [vertex] (2)    at (0,1)  {1};
    \node [vertex] (3)    at (1,1)  {1};
    \node [vertex] (4)    at (3,1)  {1};
    \node [vertex] (12)   at (-1,2) {1};
    \node [vertex] (13)   at (0,2)  {1};
    \node [vertex] (23)   at (1,2)  {0};
    \node [vertex] (14)   at (2,2)  {1};
    \node [vertex] (24)   at (3,2)  {1};
    \node [vertex] (34)   at (4,2)  {1};
    \node [vertex] (123)  at (0,3)  {0};
    \node [vertex] (124)  at (2,3)  {1};
    \node [vertex] (134)  at (3,3)  {1};
    \node [vertex] (234)  at (4,3)  {0};
    \node [vertex] (1234) at (3,4)  {0};
    \node [rred] (legend) at (1.5,4) {$g_1$};
    \draw \hasselines ;
    \draw[ultra thick, ->] (0) -- (1);
    \draw[ultra thick, ->] (2) -- (12);
    \draw[ultra thick, ->] (3) -- (13);
    \draw[ultra thick, ->] (4) -- (14);
    \draw[ultra thick, ->] (23) -- (123);
    \draw[ultra thick, ->] (24) -- (124);
    \draw[ultra thick, ->] (34) -- (134);
    \draw[ultra thick, ->] (234) -- (1234);
  \end{tikzpicture}
 \]
\[
  \begin{tikzpicture}[yscale=.6,xscale=.7]
    \node [vertex] (0)    at (0,0)  {0};
    \node [vertex] (1)    at (-1,1) {1};
    \node [vertex] (2)    at (0,1)  {1};
    \node [vertex] (3)    at (1,1)  {1};
    \node [vertex] (4)    at (3,1)  {1};
    \node [vertex] (12)   at (-1,2) {2};
    \node [vertex] (13)   at (0,2)  {1};
    \node [vertex] (23)   at (1,2)  {1};
    \node [vertex] (14)   at (2,2)  {1};
    \node [vertex] (24)   at (3,2)  {2};
    \node [vertex] (34)   at (4,2)  {1};
    \node [vertex] (123)  at (0,3)  {1};
    \node [vertex] (124)  at (2,3)  {2};
    \node [vertex] (134)  at (3,3)  {1};
    \node [vertex] (234)  at (4,3)  {1};
    \node [vertex] (1234) at (3,4)  {1};
    \node [rred] (legend) at (1.5,4) {$g_2$};
    \draw \hasselines ;
    \draw[ultra thick, ->] (0) -- (2);
    \draw[ultra thick, ->] (1) -- (12);
    \draw[ultra thick, ->] (3) -- (23);
    \draw[ultra thick, ->] (4) -- (24);
    \draw[ultra thick, ->] (13) -- (123);
    \draw[ultra thick, ->] (14) -- (124);
    \draw[ultra thick, ->] (34) -- (234);
    \draw[ultra thick, ->] (134) -- (1234);
  \end{tikzpicture}
\,
\begin{tikzpicture}[yscale=.6,xscale=.7]
    \node [vertex] (0)    at (0,0)  {0};
    \node [vertex] (1)    at (-1,1) {1};
    \node [vertex] (2)    at (0,1)  {1};
    \node [vertex] (3)    at (1,1)  {1};
    \node [vertex] (4)    at (3,1)  {1};
    \node [vertex] (12)   at (-1,2) {2};
    \node [vertex] (13)   at (0,2)  {2};
    \node [vertex] (23)   at (1,2)  {2};
    \node [vertex] (14)   at (2,2)  {1};
    \node [vertex] (24)   at (3,2)  {2};
    \node [vertex] (34)   at (4,2)  {2};
    \node [vertex] (123)  at (0,3)  {3};
    \node [vertex] (124)  at (2,3)  {2};
    \node [vertex] (134)  at (3,3)  {2};
    \node [vertex] (234)  at (4,3)  {3};
    \node [vertex] (1234) at (3,4)  {3};
    \node [rred] (legend) at (1.5,4) {$g_3$};
    \draw \hasselines ;
    \draw[ultra thick, ->] (0) -- (3);
    \draw[ultra thick, ->] (1) -- (13);
    \draw[ultra thick, ->] (2) -- (23);
    \draw[ultra thick, ->] (4) -- (34);
    \draw[ultra thick, ->] (12) -- (123);
    \draw[ultra thick, ->] (14) -- (134);
    \draw[ultra thick, ->] (24) -- (234);
    \draw[ultra thick, ->] (124) -- (1234);
  \end{tikzpicture}
  \,
  \begin{tikzpicture}[yscale=.6,xscale=.7]
    \node [vertex] (0)    at (0,0)  {0};
    \node [vertex] (1)    at (-1,1) {1};
    \node [vertex] (2)    at (0,1)  {1};
    \node [vertex] (3)    at (1,1)  {1};
    \node [vertex] (4)    at (3,1)  {1};
    \node [vertex] (12)   at (-1,2) {2};
    \node [vertex] (13)   at (0,2)  {2};
    \node [vertex] (23)   at (1,2)  {2};
    \node [vertex] (14)   at (2,2)  {2};
    \node [vertex] (24)   at (3,2)  {3};
    \node [vertex] (34)   at (4,2)  {3};
    \node [vertex] (123)  at (0,3)  {3};
    \node [vertex] (124)  at (2,3)  {4};
    \node [vertex] (134)  at (3,3)  {4};
    \node [vertex] (234)  at (4,3)  {5};
    \node [vertex] (1234) at (3,4)  {6};
    \node [rred] (legend) at (1,4) {$g=g_4$};
    \draw \hasselines ;
    \draw[ultra thick, ->] (0) -- (4);
    \draw[ultra thick, ->] (1) -- (14);
    \draw[ultra thick, ->] (2) -- (24);
    \draw[ultra thick, ->] (3) -- (34);
    \draw[ultra thick, ->] (12) -- (124);
    \draw[ultra thick, ->] (13) -- (134);
    \draw[ultra thick, ->] (23) -- (234);
    \draw[ultra thick, ->] (123) -- (1234);
  \end{tikzpicture}
  \]
\caption{Yates's algorithm on the indicator function of the nonempty
  independent sets of the graph $G$. %
  Arrows indicate how the value of $g_i(S)$ for $i\in S$ is computed by
  adding $g_{i-1}(S\setminus\{i\})$ to the `previous' value $g_{i-1}(S)$.
}
\end{figure}

\begin{pf}[Proof of \eqref{eq: yates}]
  The intuition is that $g_0(S),\ldots,g_n(S)$ approach $g(S)$
  `coordinate-wise' by fixing fewer and fewer bits of $S$. %
  To be precise, for $i=1,\ldots, n$,
  \begin{equation}\label{eq: Y ind}
    g_i(S)= \sum_{R\subseteq S} [S\cap \{i+1,\ldots,n\} =
  R\cap \{i+1,\ldots, n\} ] \cdot f(R)\,.
\end{equation}
In particular, $g_n(S)=\sum_{R\subseteq S} f(R)$. %
Correctness of \eqref{eq: yates} is established by a straightforward
but tedious induction argument for \eqref{eq: Y ind}. %
The base case $g_0=f$ is immediate. %
For the inductive step, adopt the notation $S(i)$ for $S\cap
\{i+1,\ldots, n\}$. %
Then the right hand side of \eqref{eq: Y ind} can be written as
  \begin{align*}
    \sum_{R\subseteq S} [S(i) = R(i)] f(R) & \\
=     \sum_{\substack{R\subseteq S\\i\in R}} [S(i)= R(i)] f(R) &+
    \sum_{\substack{R\subseteq S\\i\notin R}} [S(i)= R(i)] f(R)\,.
  \end{align*}
  If $i\notin S$ then the first sum vanishes and the second sum
  simplifies to 
  \[
  \sum_{R\subseteq S} [S(i-1)= R(i-1)] f(R)  = g_{i-1}(S)
  \]
  by induction. %
  If $i\in S$ then we can rewrite both sums to
  \begin{align*}
    \sum_{R\subseteq S} [S(i-1)= R(i-1)] f(R) &+
    \sum_{\substack{R\subseteq S\\i\notin S}} [S(i-1)= R(i-1)] f(R)\\
    = g_{i-1}(S) &+ g_{i-1}(S\setminus\{i\})
  \end{align*}
  by induction. %
  Finally, by \eqref{eq: yates} the entire expression equals $g_i(S)$.
\end{pf}

\paragraph{Perspective.}
As before, our approach is basically dynamic programming for a
decrease-and-conquer recurrence. %
The time and space requirements are within a linear factor of the ones
given in \S\ref{sec: colouring}. %
However, the expression \eqref{eq: yates} is more general and does not
depend on the structure of independent sets. %
It applies to any function $f\colon 2^N\rightarrow R$ from subsets to a
ring, extending the algorithm to many other covering problems than
graph colouring. %

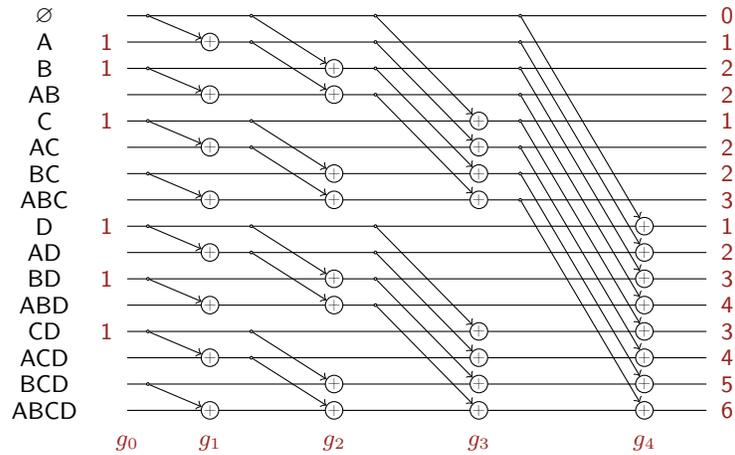
\begin{figure}
  \tikzstyle{gate}=[fill=white,draw,circle, inner sep=0pt,font=\tiny]
  \tikzstyle{setlabel}=[font=\sf\footnotesize]
\[  \begin{tikzpicture}[yscale=.35, xscale=1.1]
    \node [setlabel] (0) at (0,15) {$\emptyset$};
    \node [setlabel] (A) at (0,14) {A};
    \node [setlabel] (B) at (0,13) {B};
    \node [setlabel] (AB) at (0,12) {AB};
    \node [setlabel] (C) at (0,11) {C};
    \node [setlabel] (AC) at (0,10) {AC};
    \node [setlabel] (BC) at (0,9) {BC};
    \node [setlabel] (ABC) at (0,8) {ABC};
    \node [setlabel] (D) at (0,7)   {D};
    \node [setlabel] (AD) at (0,6) {AD};
    \node [setlabel] (BD) at (0,5) {BD};
    \node [setlabel] (ABD) at (0,4) {ABD};
    \node [setlabel] (CD) at (0,3) {CD};
    \node [setlabel] (ACD) at (0,2) {ACD};
    \node [setlabel] (BCD) at (0,1) {BCD};
    \node [setlabel] (ABCD) at (0,0) {ABCD};
    \foreach \i in {A,B,C,D,BD,CD}
    \node [setlabel, rred] at ($(\i)+(.75,0)$) {1};
    \foreach \i in {0,...,15}
    \draw (0,\i)+(1,0) -- (8,\i);
    \foreach \i in {0, B, C, D,BC, BD, CD, BCD}
    { 
      \node [gate] (from) at ($(\i)+(1.25,0)$) {};
      \node [gate] (to)   at ($(from)+(.75,-1)$) {+};
      \draw [->]          (from) -- (to);
    }
    \foreach \i in {0, A, C, D,AC, AD, CD, ACD}
    {
      \node [gate] (from) at ($(\i)+(2.5,0)$) {};
      \node [gate] (to)   at ($(from)+(1,-2)$) {+};
      \draw [->]          (from) -- (to);      
    }
    \foreach \i in {0, A, B, D,AB, AD, BD, ABD}
    {
      \node [gate] (from) at ($(\i)+(4,0)$) {};
      \node [gate] (to)   at ($(from)+(1.25,-4)$) {+};
      \draw [->]          (from) -- (to);
    }
    \foreach \i in {0, A, B, C,AB, AC, BC, ABC}
    {
      \node [gate] (from) at ($(\i)+(5.75,0)$) {};
      \node [gate] (to)   at ($(from)+(1.5,-8)$) {+};
      \draw [->]          (from) -- (to);
    }
    \node [rred] at (1,-1.25) {$g_0$};
    \node [rred] at (2,-1.25) {$g_1$};
    \node [rred] at (3.5,-1.25) {$g_2$};
    \node [rred] at (5.25,-1.25) {$g_3$};
    \node [rred] at (7.25,-1.25) {$g_4$};
    \node [setlabel,rred] at ($ (0) + (8.25,0) $)    {0};
    \node [setlabel,rred] at ($ (A) + (8.25,0) $)    {1};
    \node [setlabel,rred] at ($ (B) + (8.25,0) $)    {2};
    \node [setlabel,rred] at ($ (AB) + (8.25,0) $)   {2};
    \node [setlabel,rred] at ($ (C) + (8.25,0) $)    {1};
    \node [setlabel,rred] at ($ (AC) + (8.25,0) $)   {2};
    \node [setlabel,rred] at ($ (BC) + (8.25,0) $)   {2};
    \node [setlabel,rred] at ($ (ABC) + (8.25,0) $)  {3};
    \node [setlabel,rred] at ($ (D) + (8.25,0) $)    {1};
    \node [setlabel,rred] at ($ (AD) + (8.25,0) $)   {2};
    \node [setlabel,rred] at ($ (BD) + (8.25,0) $)   {3};
    \node [setlabel,rred] at ($ (ABD) + (8.25,0) $)  {4};
    \node [setlabel,rred] at ($ (CD) + (8.25,0) $)   {3};
    \node [setlabel,rred] at ($ (ACD) + (8.25,0) $)  {4};
    \node [setlabel,rred] at ($ (BCD) + (8.25,0) $)  {5};
    \node [setlabel,rred] at ($ (ABCD) + (8.25,0) $) {6};
  \end{tikzpicture}\]
  \caption{\label{fig: network}
    Yates's algorithm for the zeta transform.}
\end{figure}

Yates's algorithm has much in common with the fast Fourier
transform. %
We can illustrate its operation using a butterfly-like network, see
Fig.~\ref{fig: network}.

Here we used Yates's algorithm to compute \eqref{eq: ind as zeta}, but
the method is more general than that. %
For example, it computes the \emph{M\"obius transform}, see \eqref{eq:
  moebius transform} below, and many others. %
A classical treatment of the algorithm appears in \cite{Knuth:1998}, recent
applications and modifications are in \cite{BHKKTA} and the
forthcoming journal version of \cite{BHKK07}.

\section{M\"obius inversion.}
\label{sec: mobin}

Let $f\colon 2^N\rightarrow \{0,1\}$ be a function of subsets of $N$ to
$\{0,1\}$ (indeed, any ring would do). %
To connect to the graph colouring example from \S\ref{sec: colouring},
think of $f$ as the indicator function of the nonempty independent sets in a
graph. %
The \emph{zeta transform} of $f$ is the function $(f\zeta)\colon
2^N\rightarrow \{0,1\}$ defined point-wise by
\begin{equation}
\label{eq: zeta transform} (f\zeta)(T)= \sum_{S\subseteq T} f(S)\,.
\end{equation}
The brackets around $(f\zeta)$ are usually omitted. %
The \emph{M\"obius transform} of $f$ is the function $(f\mu)\colon
2^N\rightarrow \{0,1\}$ defined point-wise by
\begin{equation}
\label{eq: moebius transform}
(f\mu)(T)= \sum_{S\subseteq T} (-1)^{|T\setminus S|} f(S),
\end{equation}

This allows us to state the principle of inclusion--exclusion in yet
another way:
\begin{equation}\label{eq: pie mobin}
 f  \zeta \mu = f \mu \zeta  = f\,.
\end{equation}

\begin{pf}
  We show $f\zeta \mu= f$, the other argument is similar.
  \begin{equation*}
    \begin{split}
    f\zeta \mu(T)
    &=  \sum_{S\subseteq T} (-1)^{|T\setminus S|} \sum_{R\subseteq S} f(R) \\
    &= \sum_S \sum_R [S\subseteq T] [R\subseteq S]
    (-1)^{|T\setminus S|} f(R) \\
    &=\sum_Rf(R)\sum_S [R\subseteq S\subseteq T] (-1)^{|T\setminus S|}\,.\\
    \end{split}
  \end{equation*}
  By \eqref{eq: pie}, the inner sum equals $[R=T]$, so the 
  expression simplifies to $f(T)$.
\end{pf}

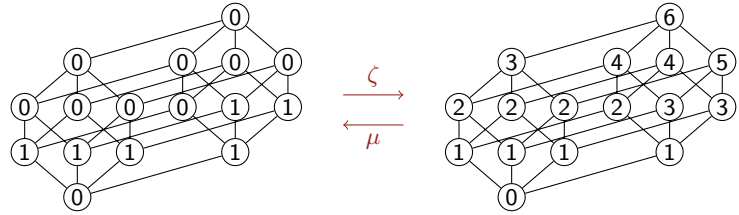
\begin{figure}
  \[
 \vcenter{\hbox{
     \begin{tikzpicture}[yscale=.6,xscale=.7]
    \node [vertex] (0)    at (0,0)  {0};
    \node [vertex] (1)    at (-1,1) {1};
    \node [vertex] (2)    at (0,1)  {1};
    \node [vertex] (3)    at (1,1)  {1};
    \node [vertex] (4)    at (3,1)  {1};
    \node [vertex] (12)   at (-1,2) {0};
    \node [vertex] (13)   at (0,2)  {0};
    \node [vertex] (23)   at (1,2)  {0};
    \node [vertex] (14)   at (2,2)  {0};
    \node [vertex] (24)   at (3,2)  {1};
    \node [vertex] (34)   at (4,2)  {1};
    \node [vertex] (123)  at (0,3)  {0};
    \node [vertex] (124)  at (2,3)  {0};
    \node [vertex] (134)  at (3,3)  {0};
    \node [vertex] (234)  at (4,3)  {0};
    \node [vertex] (1234) at (3,4)  {0};
    \draw \hasselines ;
  \end{tikzpicture}
}}
\quad
\vcenter{\hbox{
  \begin{tikzpicture}[rred,scale=.4]
    \draw[->] (0,1) -- node[above] {$\zeta$} (2,1);
    \draw[->] (2,0) -- node[below]{$\mu$} (0,0);
  \end{tikzpicture}
}}
  \quad
  \vcenter{\hbox{
  \begin{tikzpicture}[yscale=.6,xscale=.7]
    \node [vertex] (0)    at (0,0)  {0};
    \node [vertex] (1)    at (-1,1) {1};
    \node [vertex] (2)    at (0,1)  {1};
    \node [vertex] (3)    at (1,1)  {1};
    \node [vertex] (4)    at (3,1)  {1};
    \node [vertex] (12)   at (-1,2) {2};
    \node [vertex] (13)   at (0,2)  {2};
    \node [vertex] (23)   at (1,2)  {2};
    \node [vertex] (14)   at (2,2)  {2};
    \node [vertex] (24)   at (3,2)  {3};
    \node [vertex] (34)   at (4,2)  {3};
    \node [vertex] (123)  at (0,3)  {3};
    \node [vertex] (124)  at (2,3)  {4};
    \node [vertex] (134)  at (3,3)  {4};
    \node [vertex] (234)  at (4,3)  {5};
    \node [vertex] (1234) at (3,4)  {6};
    \draw \hasselines ;
  \end{tikzpicture}
}}
  \]
\caption{M\"obius inversion.}
\end{figure}

\paragraph{Perspective.} 
For completeness, let us also derive \eqref{eq: pie} from \eqref{eq:
  pie mobin}, to see that the two claims are equivalent. %
Consider two sets $R$ and $T$. %
Define $f(Q)=[Q= R]$. %
Then, expanding \eqref{eq: pie mobin},
\[ 
\begin{split}
[R=T] & =  f(T) = \sum_{S\subseteq T} (-1)^{|T\setminus S|}
\sum_{Q\subseteq S} f(Q) = \sum_{S\subseteq T} (-1)^{|T\setminus S|}
[R\subseteq S] \\
& =
 \sum_{R\subseteq S\subseteq T} (-1)^{|T\setminus S|}\,.
\end{split}
\]

\section{Covering by M\"obius inversion.}
We now give alternative argument for the graph colouring expression
\eqref{eq: ie c}.

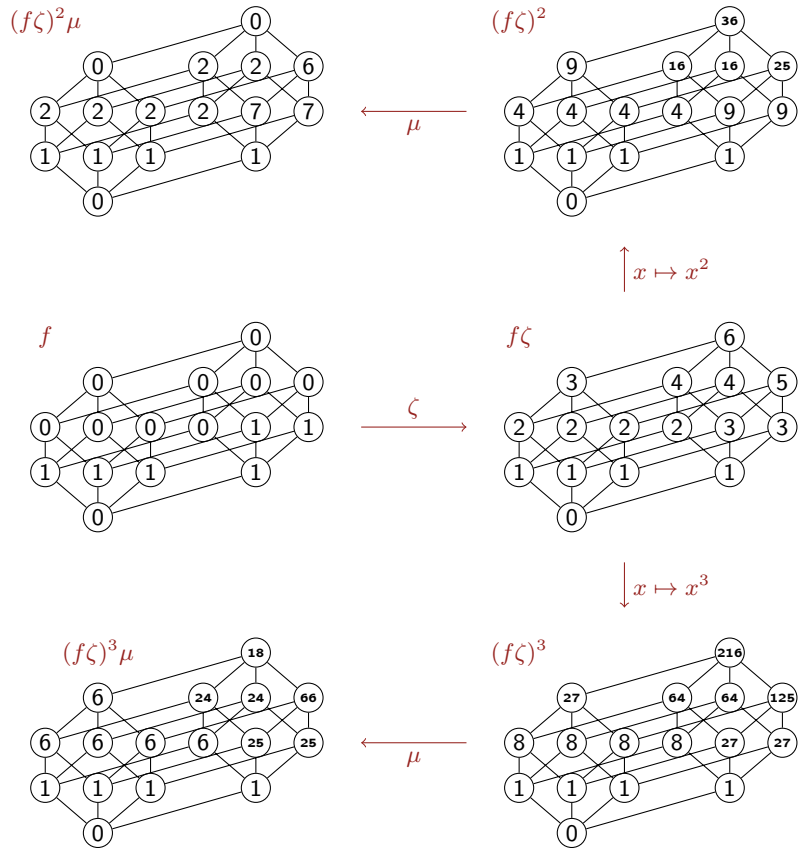
\begin{figure}[t]
  \tikzstyle{vvertex}=[inner sep= 0pt,minimum size=11pt,draw,circle,font=\sf\small]
  \[\begin{tikzpicture}[yscale=.6,xscale=.7]
     \begin{scope}
    \node [vvertex] (0)    at (0,0)  {0};
    \node [vvertex] (1)    at (-1,1) {1};
    \node [vvertex] (2)    at (0,1)  {1};
    \node [vvertex] (3)    at (1,1)  {1};
    \node [vvertex] (4)    at (3,1)  {1};
    \node [vvertex] (12)   at (-1,2) {0};
    \node [vvertex] (13)   at (0,2)  {0};
    \node [vvertex] (23)   at (1,2)  {0};
    \node [vvertex] (14)   at (2,2)  {0};
    \node [vvertex] (24)   at (3,2)  {1};
    \node [vvertex] (34)   at (4,2)  {1};
    \node [vvertex] (123)  at (0,3)  {0};
    \node [vvertex] (124)  at (2,3)  {0};
    \node [vvertex] (134)  at (3,3)  {0};
    \node [vvertex] (234)  at (4,3)  {0};
    \node [vvertex] (1234) at (3,4)  {0};
    \draw \hasselines ;
    \node [rred] at (-1,4) {$f$};
  \end{scope}
  \begin{scope}[xshift=9cm]
    \node [vvertex] (0)    at (0,0)  {0};
    \node [vvertex] (1)    at (-1,1) {1};
    \node [vvertex] (2)    at (0,1)  {1};
    \node [vvertex] (3)    at (1,1)  {1};
    \node [vvertex] (4)    at (3,1)  {1};
    \node [vvertex] (12)   at (-1,2) {2};
    \node [vvertex] (13)   at (0,2)  {2};
    \node [vvertex] (23)   at (1,2)  {2};
    \node [vvertex] (14)   at (2,2)  {2};
    \node [vvertex] (24)   at (3,2)  {3};
    \node [vvertex] (34)   at (4,2)  {3};
    \node [vvertex] (123)  at (0,3)  {3};
    \node [vvertex] (124)  at (2,3)  {4};
    \node [vvertex] (134)  at (3,3)  {4};
    \node [vvertex] (234)  at (4,3)  {5};
    \node [vvertex] (1234) at (3,4)  {6};
    \draw \hasselines ;
    \node [rred] at (-1,4) {$f\zeta$};
    \end{scope}
    \begin{scope}[shift={(9cm,7cm)}]
      \node [vvertex] (0)    at (0,0)  {0};
      \node [vvertex] (1)    at (-1,1) {1};
      \node [vvertex] (2)    at (0,1)  {1};
      \node [vvertex] (3)    at (1,1)  {1};
      \node [vvertex] (4)    at (3,1)  {1};
      \node [vvertex] (12)   at (-1,2) {4};
      \node [vvertex] (13)   at (0,2)  {4};
      \node [vvertex] (23)   at (1,2)  {4};
      \node [vvertex] (14)   at (2,2)  {4};
      \node [vvertex] (24)   at (3,2)  {9};
      \node [vvertex] (34)   at (4,2)  {9};
      \node [vvertex] (123)  at (0,3)  {9};
      \node [vvertex] (124)  at (2,3)  {\tiny 16};
      \node [vvertex] (134)  at (3,3)  {\tiny 16};
      \node [vvertex] (234)  at (4,3)  {\tiny 25};
      \node [vvertex] (1234) at (3,4)  {\tiny 36};
      \draw \hasselines ;
      \node [rred] at (-1,4) {$(f\zeta)^2$};
    \end{scope}
    \begin{scope}[shift={(9cm,-7cm)}]
      \node [vvertex] (0)    at (0,0)  {0};
      \node [vvertex] (1)    at (-1,1) {1};
      \node [vvertex] (2)    at (0,1)  {1};
      \node [vvertex] (3)    at (1,1)  {1};
      \node [vvertex] (4)    at (3,1)  {1};
      \node [vvertex] (12)   at (-1,2) {8};
      \node [vvertex] (13)   at (0,2)  {8};
      \node [vvertex] (23)   at (1,2)  {8};
      \node [vvertex] (14)   at (2,2)  {8};
      \node [vvertex] (24)   at (3,2)  {\tiny 27};
      \node [vvertex] (34)   at (4,2)  {\tiny 27};
      \node [vvertex] (123)  at (0,3)  {\tiny 27};
      \node [vvertex] (124)  at (2,3)  {\tiny 64};
      \node [vvertex] (134)  at (3,3)  {\tiny 64};
      \node [vvertex] (234)  at (4,3)  {\tiny 125};
      \node [vvertex] (1234) at (3,4)  {\tiny 216};
      \draw \hasselines ;
      \node [rred] at (-1,4) {$(f\zeta)^3$};
    \end{scope}
    \begin{scope}[shift={(0cm,7cm)}]
      \node [vvertex] (0)    at (0,0)  {0};
      \node [vvertex] (1)    at (-1,1) {1};
      \node [vvertex] (2)    at (0,1)  {1};
      \node [vvertex] (3)    at (1,1)  {1};
      \node [vvertex] (4)    at (3,1)  {1};
      \node [vvertex] (12)   at (-1,2) {2};
      \node [vvertex] (13)   at (0,2)  {2};
      \node [vvertex] (23)   at (1,2)  {2};
      \node [vvertex] (14)   at (2,2)  {2};
      \node [vvertex] (24)   at (3,2)  {7};
      \node [vvertex] (34)   at (4,2)  {7};
      \node [vvertex] (123)  at (0,3)  {0};
      \node [vvertex] (124)  at (2,3)  {2};
      \node [vvertex] (134)  at (3,3)  {2};
      \node [vvertex] (234)  at (4,3)  {6};
      \node [vvertex] (1234) at (3,4)  {0};
      \draw \hasselines ;
      \node [rred] at (-1,4) {$(f\zeta)^2\mu$};
    \end{scope}
    \begin{scope}[shift={(0cm,-7cm)}]
      \node [vvertex] (0)    at (0,0)  {0};
      \node [vvertex] (1)    at (-1,1) {1};
      \node [vvertex] (2)    at (0,1)  {1};
      \node [vvertex] (3)    at (1,1)  {1};
      \node [vvertex] (4)    at (3,1)  {1};
      \node [vvertex] (12)   at (-1,2) {6};
      \node [vvertex] (13)   at (0,2)  {6};
      \node [vvertex] (23)   at (1,2)  {6};
      \node [vvertex] (14)   at (2,2)  {6};
      \node [vvertex] (24)   at (3,2)  {\tiny 25};
      \node [vvertex] (34)   at (4,2)  {\tiny 25};
      \node [vvertex] (123)  at (0,3)  {6};
      \node [vvertex] (124)  at (2,3)  {\tiny 24};
      \node [vvertex] (134)  at (3,3)  {\tiny 24};
      \node [vvertex] (234)  at (4,3)  {\tiny 66};
      \node [vvertex] (1234) at (3,4)  {\tiny 18};
      \draw \hasselines ;
      \node [rred] at (0,4) {$(f\zeta)^3\mu$};
    \end{scope}
    \draw[rred,->] (5,2)  -- node [above] {$\zeta$} (7,2);
    \draw[rred,->] (7,9) -- node [below] {$\mu$}  (5,9);
    \draw[rred,->] (7,-5) -- node [below] {$\mu$}  (5,-5);
    \draw[rred,->] (10,5) -- node [right] {$x\mapsto x^2$}  (10,6);
    \draw[rred,->] (10,-1) -- node [right] {$x\mapsto x^3$}  (10,-2);
 \end{tikzpicture}\]
  \caption{\label{fig: mobin}Covering by M\"obius inversion 
    for $k=2$ and $k=3$. }
\end{figure}

Let $f\colon 2^N\rightarrow \{0,1\}$ be the indicator function of the
nonempty independent sets of a graph $G=(N,E)$. %
We want to count the number of ways so cover $N$ with $k$ nonempty independent
sets. %
Define $g(S)$ to be the number of ways to choose $k$ nonempty independent sets
whose union is $S$. %
Then we claim 
\[
g\zeta=(f\zeta)^k\,.
\] 
To see this, for every $T\subseteq V$, view $g\zeta(T)$ and
$\bigl(f\zeta (T)\bigr)^k$ as two different ways of counting the
number of ways so select $k$ nonempty independent subsets of $T$.  %
Now, by M\"obius inversion \eqref{eq: pie mobin}, we
have \[g=(f\zeta)^k\mu\,,\] which is the left hand side of \eqref{eq:
  ie c}. %
In fact, we can now rewrite and understand the left hand side of
\eqref{eq: ie c} as
\[
\begin{tikzpicture}[start chain, node distance = 0pt,text height=1.5ex,text depth=.25ex]
  \node (mu) [on chain] {$\displaystyle \sum_{S\subseteq N} (-1)^{|N\setminus
      S|}$};
  \node [on chain] {$\displaystyle\biggl($} ; 
  \node (zeta) [on chain] {$\displaystyle \sum_{R\subseteq S}$};
  \node (f) [draw=rred, on chain] {$f(R)$};
  \node [on chain] {$\displaystyle \biggr)$};
  \node (op) [draw=rred,on chain,yshift=2ex] {$\scriptstyle k$};
  \draw [rred,decorate,decoration={brace,mirror,raise=3ex}] 
        (mu.south west) -- node[below=3ex]{$\mu$} (mu.south east);
  \draw [rred,decorate,decoration={brace,mirror,raise=3ex}] 
        (zeta.south west) -- node[below=3ex]{$\zeta$} (zeta.south east);
  \node [rred, anchor=west] (lbl1) at ($(op) + (1,-.75)$)
  {\emph{\small function in the original
      domain}};
  \draw [rred, ->] (lbl1.west) -- (f);
  \node [rred,anchor=west] (lbl2) at ($(op) + (1,-.25)$) {\emph{\small operation in the
      transformed domain}};
  \draw [rred, ->] (lbl2.west) -- (op);
\end{tikzpicture}
\]

\paragraph{Perspective.} The fact that $f$ was the indicator function
of the \emph{independent sets} played no role in this argument. %
It works for a many covering problems, and with some work also for
packing and partitioning problems.

\medskip Taxonomically, we can view inclusion--exclusion as a
\emph{transform-and-conquer} technique, like the Fourier transform. %
This can be expressed succinctly in terms of M\"obius inversion,
illustrated in Fig.~\ref{fig: mobin}. %
The zeta transform translates the original problem into a different
domain, where the computation is often easier, and the M\"obius
transform translates back. %
In the covering example, the operation in the transformed domain,
exponentiation, is particularly simple. %
The idea extends to many other operations in the transformed domain,
see \cite{BHKK07}.

\section*{Concluding remarks}
A comprehensive presentation of many of the ideas mentioned here
appears in a recent monograph \cite{FK}, with many additional
examples. %
In particular a whole chapter is devoted to subset convolution, the
most glaring omission from the present introduction.

I owe much of my understanding of this topic to illuminating
conversations with my collaborators, Andreas Bj\"orklund, Petteri
Kaski, and Mikko Koivisto. %

\bibliographystyle{abbrv}

\end{document}